# Efficient cathodoluminescence of monolayer transitional metal dichalcogenides in a van der Waals heterostructure

Shoujun Zheng[1], Jinkyu So[2], Fucai Liu[3], Zheng Liu[3], Nikolay Zheludev[1,*], and Hong Jin Fan[1,2,*]

[1]*Centre for Disruptive Photonic Technologies, Nanyang Technological University, 637371, Singapore*

[2]*Division of Physics and Applied Physics, School of Physical and Mathematical Sciences, Nanyang Technological University, 637371, Singapore*

[3]*School of Materials Science and Engineering, Nanyang Technological University, 639798, Singapore*

S. Zheng and J. So contributed equally to this work.

*Corresponding authors. Emails: NZheludev@ntu.edu.sg (N.Z.); fanhj@ntu.edu.sg (H.J.F.)

**Abstract:** Monolayer two-dimensional transitional metal dichalcogenides, such as $MoS_2$, $WS_2$ and $WSe_2$, are direct band gap semiconductors with large exciton binding energy. They attract growing attentions for opto-electronic applications including solar cells, photo-detectors, light-emitting diodes and photo-transistors, capacitive energy storage, photodynamic cancer therapy and sensing on flexible platforms. While light-induced luminescence has been widely studied, luminescence induced by injection of free electrons could promise another important applications of these new materials. However, cathodoluminescence is inefficient due to the low cross-section of the electron-hole creating process in the monolayers. Here for the first time we show that cathodoluminescence of monolayer chalcogenide semiconductors can be evidently observed in a van der Waals heterostructure when the monolayer semiconductor is sandwiched between layers of hexagonal boron nitride (hBN) with higher energy gap. The emission intensity shows a strong dependence on the thicknesses of surrounding layers and the enhancement factor is more than 1000 folds. Strain-induced exciton peak shift in the suspended heterostructure is also investigated by the cathodoluminescence spectroscopy. Our results demonstrate that $MoS_2$, $WS_2$ and $WSe_2$ could be promising cathodoluminescent materials for applications in single-photon emitters, high-energy particle detectors, transmission electron microscope displays, surface-conduction electron-emitter and field emission display technologies.

*Keywords: transitional metal dichalcogenides; van der Waals heterostructures; cathodoluminescence; strain effect; 2D heterostructures*

**Introduction**

Two-dimensional (2D) layered semiconductors, due to their weak interlayer van der Waals bonds, can be easily thinned down to atomic thickness by mechanical[1] and chemical[2] exfoliation methods, for example, graphene[3] and hexagonal boron nitride (hBN)[4]. Monolayer transitional metal dichalcogenides with the formula of $MX_2$ (M=Mo, W; X=S, Se, Te) are a type of unique semiconductors with narrow direct band gaps, large exciton binding energies, high optoconductivity, and high photoelectrochemical activity. Moreover, due to the inversion symmetry breaking, monolayer $MX_2$ are widely employed for the study of valley polarization and spin−valley coupling[5,6]. Recently, van der Waals heterostructures, composed of different 2D materials with unique band alignment and interlayer coupling, attract growing attentions not only for fundamental new physics but also for many potential applications such as tunneling transistors[7,8,9] and light-emitting diodes[10].

Cathodoluminescence, photon emission excited by a high-energy electron beam, is widely applied in the analysis of mineral compositions[11], light emitting diodes[12,13], surface plasmon mapping[14]. Compared to photoluminescence excited by light, cathodoluminescence offers a much higher excitation energy allowing the study of wide band gap materials including diamond[15] and hexagonal boron nitride (hBN)[16,17]. Due to a small excitation hotspot cathodoluminescence has been extensively used to study nanostructures including hyper-spectral imaging of plasmonic gratings[18], nanoparticles[19], nano-antenna[20], quantum well[21,22], three-dimensional nanoscale visualization of metal-dielectric nanoresonators[23] and nanoscale light sources[24,25].

In atomic layers of $MX_2$, it is challenging to detect the interband cathodoluminescence signal as the electron-hole creation cross section is extremely small. Moreover, the spatial distribution of electron-hole pairs at the interface, which is near the point of free-electron injection, is close to a 3D spherical shape of a few microns in diameter. Only a small fraction of recombination takes place in the top 2D material and most of them happen in the supporting slab. Indeed, so far only a few reports are available on cathodoluminescence study of 2D materials, including six atomic layer thick flakes of boron nitride.[26,27,28] However, cathodoluminescence from monolayer $MX_2$ has not reported.

In this report we show that cathodoluminescence emissions from monolayer $MX_2$ ($MoS_2$, $WS_2$ and $WSe_2$) can be enhanced and efficiently detected in a van der Waals heterostructure, in which the luminescent $MX_2$ layer is sandwiched between layers of hexagonal boron nitride (hBN) with higher energy gap (see schematics in Fig. 1a). Here the hBN/$MX_2$/hBN heterostructure can effectively increase the recombination probability of electron-hole pairs in the monolayer $MX_2$ in such a way that a good fraction of the electrons and holes generated in the hBN layers diffuse to and then radiative recombine in the $MX_2$ layer, leading to significant enhancement of the emission, comparatively to an isolated layer (Fig. 1b).

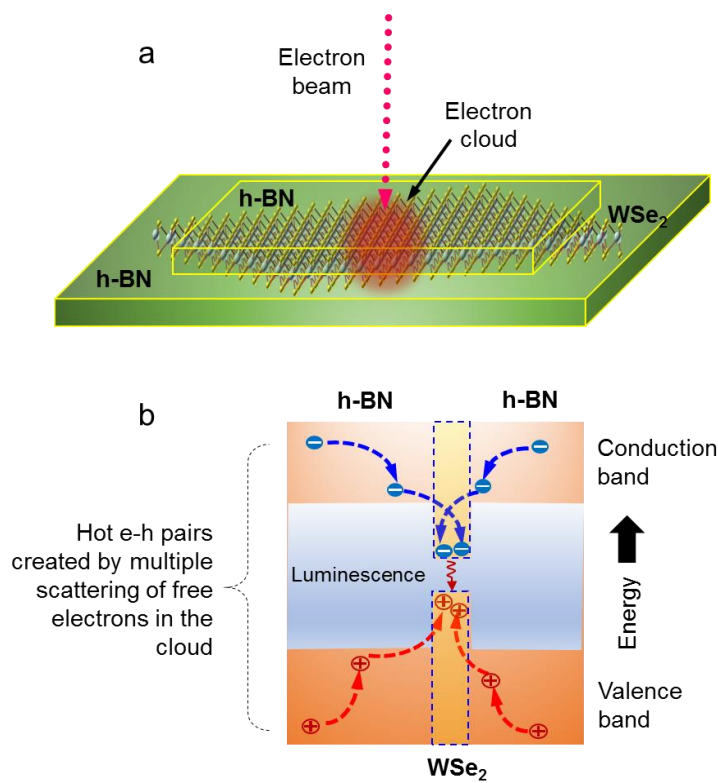

**Figure 1** | (a) Illustration of cathodoluminescence in an hBN/$MX_2$/hBN van der Waals heterostructure. (b) Process of the generation, diffusion and recombination of e-h pairs. The minor number of e-h pairs generated in the $MX_2$ layer is ignored.

**Cathodoluminescence of hBN/monolayer WSe$_2$/hBN heterostructure**

To fabricate the hBN/ $MX_2$/hBN structure, we stacked the different layers one by one by the dry pick-up transfer method[1, 29]. Firstly, an hBN flake was picked up by a polydimethylsiloxane (PDMS) film with a spincoated polyvinyl alcohol (PVA) layer. Then the hBN flake supported by the PDMS/PVA film was aligned to the monolayer WSe$_2$ flake to pick up it by the Van der Waals force (Fig. 2a). Next, the hBN/WSe$_2$ heterostructure was aligned to a large and thick bottom hBN (thickness ~100 nm). Finally the PDMS/PVA film was removed (see details in the supporting materials and Fig. S1). Figure 2b shows the optical image of the prepared hBN/WSe$_2$/hBN heterostructure. Some bubbles generated during sample transfer process and enlarged when the sample was put into SEM vacuum chamber (see in Fig. S2a), which can be proved by atomic force microscopy topography image (indicated by the bright dots in Fig. S2c). The thickness of the top hBN layer is around 4.2 nm. Furthermore, the monolayer WSe$_2$ is clearly identified by Raman mapping (Fig. S2b), in which the Raman intensities of the vibration mode A$_{1g}$ are less affected by the top hBN layer.

When cathodoluminescence measurement was done to monolayer MX$_2$ on Si substrates or freestanding monolayer MX$_2$, the emissions are too weak to be observable. However, strong emissions from the monolayer WSe$_2$ can be observed in the hBN/WSe$_2$/hBN van der Waals heterostructure (acceleration voltage of 5 keV, beam current of 36.2 nA). Cathodoluminescence mapping clearly shows the giant enhancement of the emission intensity in the hBN/WSe$_2$/hBN region (indicated by red color in Fig. 2c). The cathodoluminescence signal was only present in the hBN/WSe$_2$/hBN region (indicated by point 1), but absent in the WSe$_2$ region without the top hBN layer (point 2). Therefore, both the top and bottom hBN layer are key factors to cathodoluminescence emission of monolayer WSe$_2$. The cathodoluminescence emission peak around 1.572 eV corresponds to the excitonic energy of the monolayer WSe$_2$ (Fig. 2d). This cathodoluminescence emission peak is consistent to the photoluminescence (PL) emission peak, but with a small redshift of 16.8 meV. The disalignment between CL and PL peak position may be due to the local heating effect by the e-beam, which is consistent to the well-known temperature-induced semiconductor band gap shrinkage[30]. Similar redshifts were also observed from other MX$_2$ samples in the heterostructure. Moreover, the cathodoluminescence intensity is proportional to the electron beam currents (see Fig. S2e and inset) and unresolvable when the bean current is below 1.9 nA.

We also investigated the time-resolved cathodoluminescence by a streak camera at room temperature. The time-resolved spectrum (inset of Fig. 2d) was well fitted with a double exponential decay function of $Ae^{-t/\tau_1} + Be^{-t/\tau_2}$. The decay times of $\tau_1$ and $\tau_2$ are 36.8 and 369.9 ps, respectively. The intrinsic exciton lifetime of the monolayer MX$_2$ is predicted to be 0.1~1 ps according to theoretical calculations[31, 32]. Several experiments show that the exciton lifetime can reach several ps at the low temperature of a few K[33, 34]. However, the lifetime increases at high temperatures. The effective decay lifetime at a given temperature T is given by [33, 35] $\tau_{rad}^{eff} = \frac{3}{2}\frac{k_B T}{E_0}\tau_{rad}^0$ where $\tau_{rad}^0$ is the intrinsic exciton lifetime at 0 K, $k_B T$ is the thermal energy, $E_0$ is the kinetic energy of the exciton. According to this equation, the effective decay lifetime of the exciton becomes longer at room temperature, which even reaches hundreds ps to several ns[33]. As the lifetime of the trion is generally shorter than the one of exciton[35, 36], the decay times of 36.8 and 369.9 ps may be related to the trion and exciton, respectively.

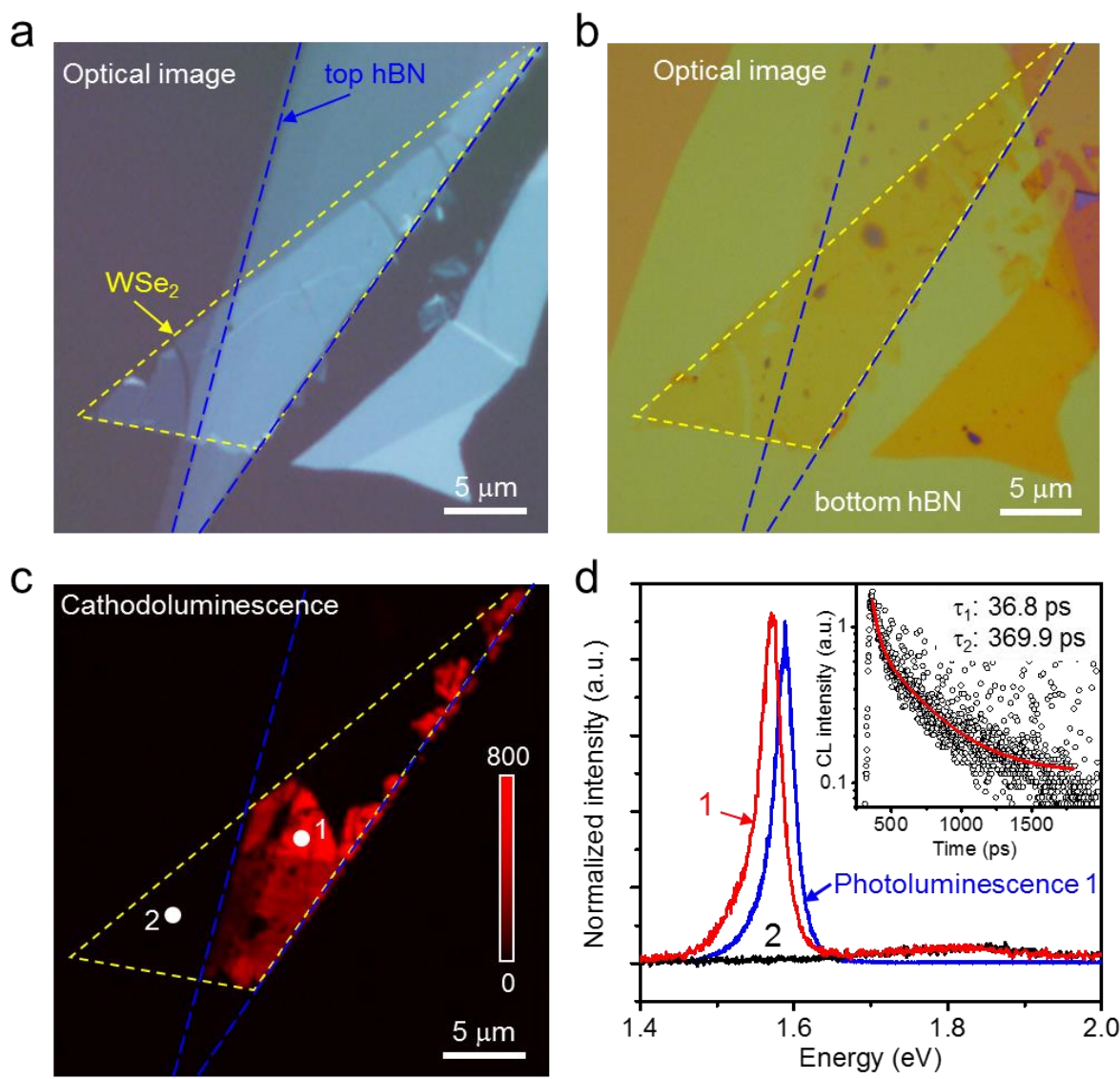

**Figure 2 | Cathodoluminescence of the monolayer WSe$_2$.** (a) Optical image of the hBN/WSe$_2$ structure on a PDMS/PVA film before the last transfer step. The top hBN and WSe$_2$ layers can be easily seen. (b) Optical image of the prepared hBN/WSe$_2$/hBN heterostructure. (c) Cathodoluminescence intensity mapping of the hBN/WSe$_2$/hBN heterostructures. The strong emission is indicated by the red color. 1: hBN/WSe$_2$/hBN part. 2: WSe$_2$/hBN part without the top hBN layer. (d) Cathodoluminescence and photoluminescence spectra. The numbers correspond to the two points in (c). Inset is the time-resolved cathodoluminescence spectrum recorded at room temperature.

**The dependence of cathodoluminescence intensity on hBN thickness**

It is found that the emission intensity is strongly dependent on the thicknesses of both top and bottom hBN layers. We fabricated an hBN/WSe$_2$/hBN sample with a flat bottom hBN layer of 165.3 nm and a top hBN layer with different thickness regions of 3.5, 11.8, and 23.0 nm (see more details in Fig. S3a-c). The cathodoluminescence mapping shows clearly the intensity difference between the three thickness regions; highest at the region of 23.0 nm and weakest at the one of 3.5 nm. This can been clearly seen from the cathodoluminescence spectra selected from the three regions (Fig. 3a). Each spectrum is the average of 20 points selected from the corresponding region to eliminate intensity inhomogeneity. The dependence of integrated intensity on top hBN thickness is plotted in Fig. 3b, which is nearly a linear relationship. Furthermore, we found that the intensity has a similar dependence on the thickness of the bottom hBN layer. We prepared another hBN/WSe$_2$/hBN sample with a flat top hBN layer (20 nm thick), and a bottom hBN layer of four thickness regions of 12.1, 21.6, 36.7, 48.3 nm (see details in Fig. S3d-f). The cathodoluminescence mapping can also be identified with four parts corresponding to the four thickness regions. The average spectra selected from each regions also attest this intensity difference (Fig. 3c). The difference in peak positions in Fig. 3 may be related to spatial-dependent strain and/or heterostructure inhomogeneity. Similar to above where the top hBN layer thickness is varied, the plot of cathodoluminescence intensity as a function of the thickness of the bottom hBN layer shows a nearly linear relationship (Fig 3d). We have also conducted similar tests on other heterostructures and obtained similar trend. Such a strong thickness dependence concords with our notion that the e-h pairs for recombination originate mainly from the hBN layers, in which a larger thickness corresponds to a higher excitation volume. According to calculations, the diffusion lengths for electrons and holes in hBN are in the μm range,[37] so the e-h pairs generated in both top and bottom hBN layers in the heterostructure can diffuse to the middle MX$_2$ layer before recombination. Therefore, it is reasonable that cathodoluminescence intensity has a strong dependence on the thickness of the hBN layers.

The question is, why both top and bottom hBN layers are necessary for evident cathodoluminescence emission. In our configuration, strong emission from WSe$_2$ requires that the generated e-h pairs can efficiently diffuse to and be trapped at the interface between the two hBN layers. Because of the potential well, the carriers are transferred to the middle MX$_2$ which is the recombination center, leading to evident luminescence emission. In case of a single top or bottom hBN layer, the generated e-h pairs are not efficiently confined at the surface of the hBN layer even with a monolayer MX$_2$. Therefore, the strong cathodoluminescence due to WSe$_2$ band gap emission is attributed to the both increased excitation volume and efficient interface confinement of e-h pairs in the sandwich configuration. The enhancement factor is more than 1000 when the thicknesses of both top and bottom hBN layers are 19.8 and 123.9 nm (Fig. S4). Finally, it is noteworthy that the bottom hBN layer cannot be replaced by the amorphous $SiO_2$ substrate, as the latter does not provide a van der Waals contact.

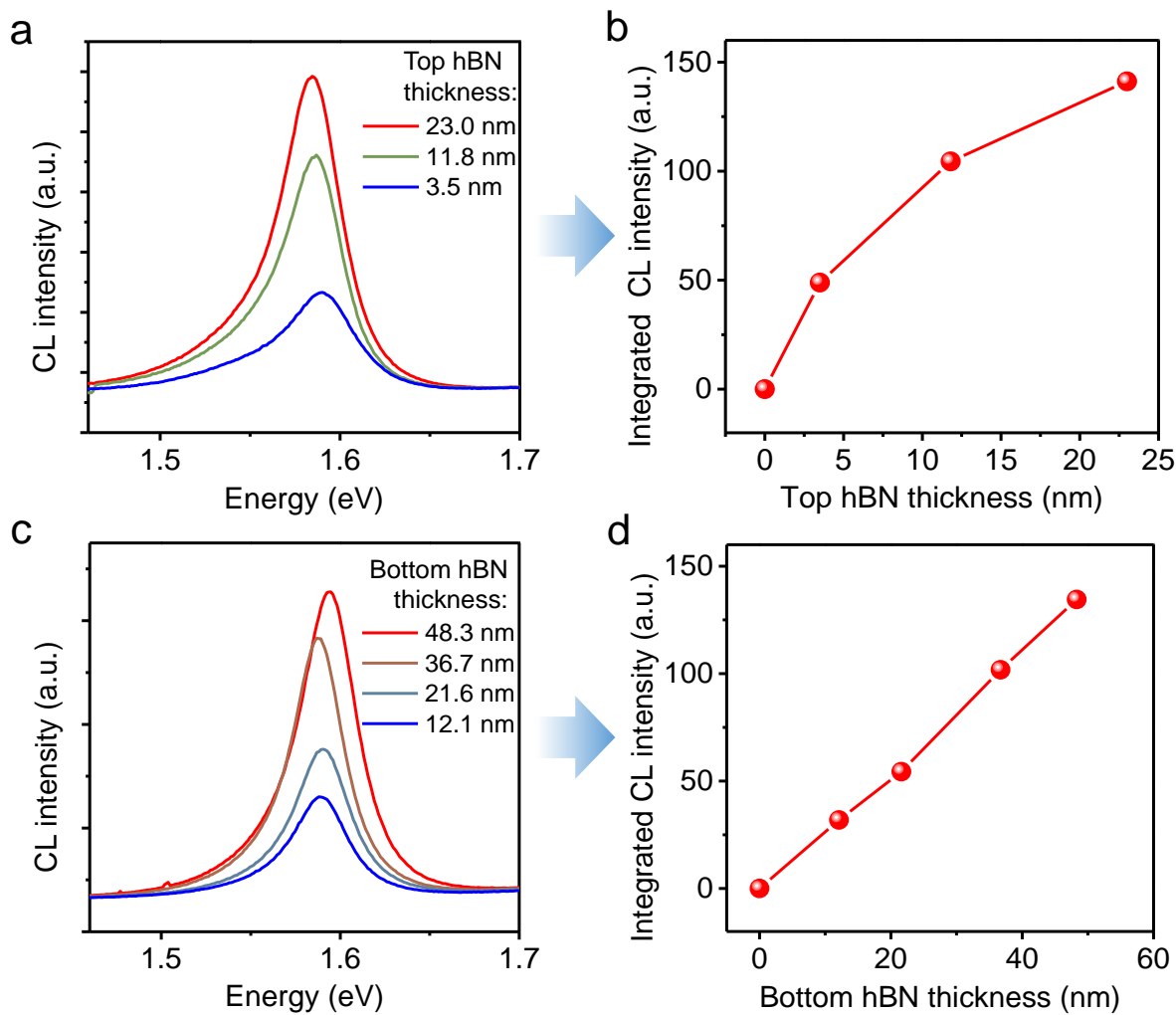

**Fig. 3 | Effects of hBN layers thickness on cathodoluminescence intensity.** (a) Spectra collected from three regions of different top hBN thicknesses. Each spectrum is the average of 20 points selected from the same thickness region. (b) The plot of integrated intensity versus the thickness of the top hBN layer. (c) Averaged spectra from the four regions of different bottom hBN thicknesses. (d) Plot of integrated intensity versus the bottom hBN layer thickness.

**Effect of strain**

Cathodoluminescence spectroscopy is also powerful in revealing spatial-resolved strain effect in 2D semiconductors. It is known that the band gap shift in $MX_2$ is sensitive to the strain, as evidenced in both theoretical calculations[38] and experiments[39, 40, 41]. We investigated the strain-induced peak shift of the monolayer $WSe_2$ by suspending the $hBN/WSe_2/hBN$ heterostructure sample on holes made on the Si substrate. Array of holes were fabricated by a focused ion beam with diameter of 1 and 2 μm, so that part of the sample is suspended (Fig. S5). Interestingly, the emission intensity of hBN/WSe$_2$/hBN suspended on the hole is much stronger than the part laid on the substrate at room temperature (Fig. 4a). By cooling down the sample, we can identify the exciton and trion peak positions clearly from the spectra. Position-dependent spectra at 10 K (Fig. 4b and 4c) clearly shows that the excitonic peak redshift with a maximum shift of 12.1 meV from the edge to the center of the hole due to strain.

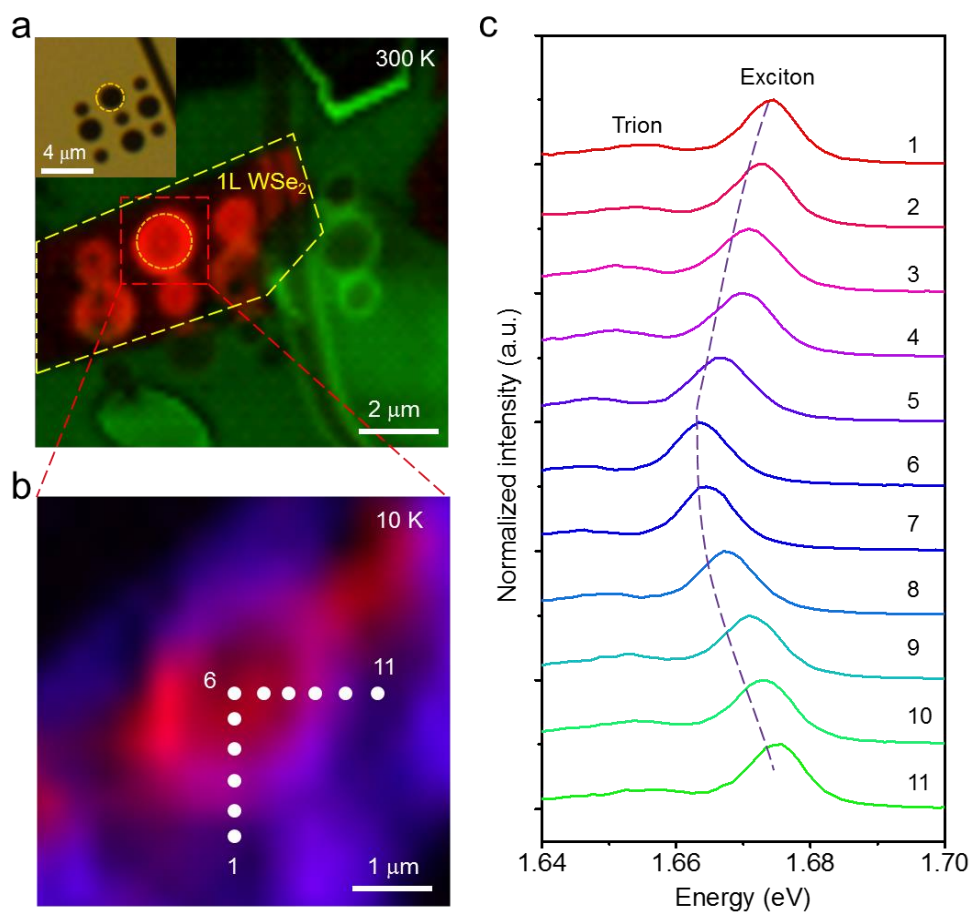

**Figure 4 | Strain effect.** (a) Cathodoluminescence mapping of the hBN/WSe$_2$/hBN sample at room temperature. Parts of the samples are suspended on focused-ion beam fabricated holes on the substrate. Inset is the Si substrate with holes before transfer. (b) Enlarged cathodoluminescence mapping at 10 K around a hole. (c) Position-dependent spectra taken from a series of points in (b).

In addition to the strain due to suspension, strain in the thin heterostructures is also inevitably introduced during the transfer process. Such inhomogeneous local strain in the heterostructures is also detectable by cathodoluminescence spectroscopy. Indeed two energy domains were observed from the hBN/WSe$_2$/hBN (indicated by purple and green colors in Fig. S6a) sample in terms of the exciton peak position at 77 K. From the CL spectra from selected points (Fig. S6b), two emission peaks can be resolved at both point A and B. The two peaks correspond to the emissions of neutral excitons and trions (charged excitons)[42]. However, the peak positions of excitons and trions at point A are 1.640 and 1.614 eV, respectively, while 1.657 and 1.623 eV at point B. The peak position difference between point A and B in the heterostructure may be caused by strain, which is possibly generated during the transfer process (e.g., bubbles).

Temperature dependent cathodoluminescence spectra of the point B were plotted in Fig. S5c. The peak positions of both excitons and trions are fitted (Fig. S5d) according to the semiempirical semiconductor band gap equation[30, 43] of $E_g(T) = E_g(0) - S\hbar\omega\left[coth\left(\frac{\hbar\omega}{2kT}\right) - 1\right]$, where $E_g(0)$ is the excitonic energy at 0 K, $S$ is a dimensionless coupling constant and $\hbar\omega$ is an average phonon energy. From the fitting curves, we extract $E_g(0)$ of the exciton and trion of the monolayer WSe$_2$ to be 1.6652 and 1.6344 eV, respectively. So, the binding energy of the trion is calculated to be 30.8 meV which is consistent with the previous report (30 meV)[32].

**Cathodoluminescence of other monolayer semiconductors**

In addition to WSe$_2$, we also performed cathodoluminescence experiments to monolayer WS$_2$ and MoS$_2$ in a van der Waals heterostructure (sandwiched by two hBN layers). The emission peak position from the WS$_2$ heterostructure locates at 1.933 eV (Fig. 5a) and that from the MoS$_2$ in heterostructure at 1.831 eV (Fig. 5b). The emission peak positions of both two heterostructures redshift with respect to their photoluminescence peak positions, similar to the case of hBN/WSe$_2$/hBN. Cathodoluminescence mapping is inhomogeneous at the MoS$_2$ sample which is most likely due to poor interface contact. Therefore, sandwiching monolayer MX$_2$ into two hBN layers is a universal approach to study cathodoluminescence emissions of monolayer MX$_2$.

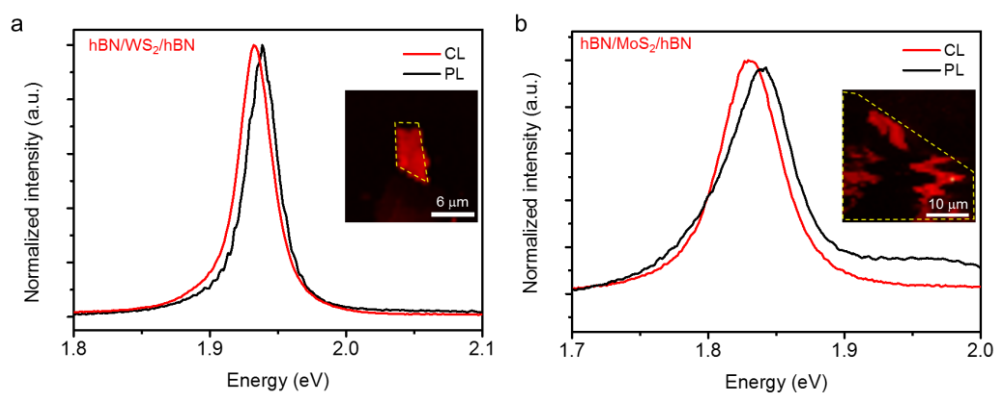

**Figure 5 | Cathodoluminescence of monolayer WS$_2$ and MoS$_2$.** (a) CL and PL spectra of the monolayer WS$_2$ in

the top-hBN (7.5 nm)/WS$_2$/bottom-hBN (299.4 nm). (b) CL and PL spectra of the monolayer MoS$_2$ in the top-hBN (13.5 nm)/MoS$_2$/bottom-hBN (168.6 nm). Insets are the corresponding cathodoluminescence intensity mappings. Yellow lines guide the shapes of monolayer TMDs in the van der Waals heterostructrues.

In summary, for the first time we have obtained evident cathodoluminescence emission from monolayer MX$_2$, including WSe$_2$, MoS$_2$ and WS$_2$, via a van der Waals configuration. In the hBN/MX$_2$/hBN heterostructure, electron beam induced e-h pairs can transfer to and be trapped in the middle MX$_2$ layer, leading to increased recombination probability within the MX$_2$ layer. Moreover, we demonstrate that cathodoluminescence sepctroscopy can be applied to study the strain-induced excitonic peak shift in monolayer MX$_2$. Because of its high spatial resolution and high beam energy, our demonstration makes cathodoluminescence spectroscopy a powerful technique to the study of 2D materials in various forms such as alloy, heterostructures or defects. The 2D monolayer-based heterostructure may promise potential applications in single-photon emitters, surface-conduction electron-emitter and field emission display technologies.

## Methods

**Fabrication of hBN/MX$_2$/hBN structures.** Monolayer MX$_2$ and hBN layers were exfoliated on Si/SiO$_2$ wafers by scotch tapes and identified by a Nikon optical microscope. Top hBN layer and MX$_2$ layer were picked up by a PDMS/PVA film one by one, then aligned and put on the bottom hBN layer. PDMS/PVA film can be removed by slow detachment which may cause loose interface contact, or immersing in water (See details in the supporting materials). The holes on the Silicon substrate were fabricated by a focused ion beam (FEI Helios NanoLab) in 1~2 μm diameter and 1 μm depth.

**AFM, Raman and photoluminescence characterization.** The thicknesses of hBN layers were measured using a Cypher ES scanning probe microscopy. Raman and PL spectra were recorded from a Witec confocal Raman spectrum with an excitation wavelength of 532 nm under 100X objective.

**Cathodoluminescence measurement**. We used a commercial Cathodoluminescence Microscope (Attolight Allalin 4027 Chronos) with an accelerating voltage of 5 kV, a beam current of 36.2 nA, and an exposure time of 50 ms for most spectra (for beam current dependent spectra, the exposure time was 10 ms). Samples were loaded into the chamber under ultra-high vacuum (<10$^{-5}$ Pa) and the emissions were collected by a UV-VIS spectrometer in the range of 180 - 1100 nm. Low-temperature spectra were collected by cooling down the sample using liquid nitrogen (78 K) and helium (10 K). For time-resolved spectra, a pulsed laser of 300 fs was used to generate electron pulse of around 5 ps and emission signals were collected by a streak camera with the range of 200 -850 nm.

**Acknowledgements**

This work is supported by Singapore Ministry of Education Academic Research Fund Tier 3 (Grant No. MOE2011-T3-1-005) and the Singapore National Research Foundation under NRF RF Award No. NRF-RF2013-08, the start-up funding from Nanyang Technological University (M4081137.070).

**Author contribution**

S. Z. conceived the experiments and prepared the samples. J. S. and S. Z. conducted the cathodoluminescence measurements. F. Liu did the Raman and photoluminescence measurements. The manuscript was written by S. Z., F. L., Z. L., N. Z., and H. J. F. H. J. Fan supervised the project. All authors participated in discussion of the data.

**Additional information**

Details of sample transfer, AFM, optical images, and more cathodoluminescence spectra.